# An Analytical Latency Model and Evaluation of the Capacity of 5G NR to Support V2X Services using V2N2V Communications

M.C. Lucas-Estañ, *Member, IEEE*, B. Coll-Perales, *Member, IEEE*, T. Shimizu, *Member, IEEE*,
J. Gozalvez, *Senior Member, IEEE*, T. Higuchi, *Member, IEEE*, S. Avedisov, *Member, IEEE*,
O. Altintas, *Member, IEEE*, M. Sepulcre, *Senior Member, IEEE*

*Abstract*— 5G has been designed to support applications such as connected and automated driving. To this aim, 5G includes a highly flexible New Radio (NR) interface that can be configured to utilize different subcarrier spacings (SCS), slot durations, scheduling, and retransmissions mechanisms. This flexibility can be exploited to support advanced V2X services with strict latency and reliability requirements using V2N2V (Vehicle-to-Network-to-Vehicles) communications instead of direct or sidelink V2V (Vehicle-to-Vehicle). To analyze this possibility, this paper presents a novel analytical model that estimates the latency of 5G at the radio network level. The model accounts for the use of different numerologies (SCS, slot durations and Cyclic Prefixes), modulation and coding schemes, full-slots or mini-slots, semi-static and dynamic scheduling, different retransmission mechanisms, and broadcast/multicast or unicast transmissions. The model has been used to first analyze the impact of different 5G NR radio configurations on the latency. We then identify which radio configurations and scenarios can 5G NR satisfy the latency and reliability requirements of V2X services using V2N2V communications. This paper considers cooperative lane changes as a case study. The results show that 5G can support advanced V2X services at the radio network level using V2N2V communications under certain conditions that depend on the radio configuration, bandwidth, service requirements and cell traffic load.

## I. INTRODUCTION

T HE 5G New Radio (NR) interface has been designed to reduce the latency and increase the reliability and spectrum efficiency of cellular networks. 5G NR offers unprecedented levels of flexibility, including the use of different numerologies, subcarrier spacings (SCS) and slot durations at the RAN (Radio Access Network)[1]. 5G NR enables dynamic (or grant-based) and semi-static (or grant-free) scheduling schemes, and the use of Hybrid-Automatic Repeat reQuest (HARQ) retransmissions or the transmission of *k* replicas of a packet for higher flexibility to meet latency and reliability demands.

The capabilities introduced by 5G NR have raised expectations on the possibility to use the 5G Uu interface to support advanced V2X (Vehicle to Everything) services using V2N (Vehicle-to-Network) or V2N2V (Vehicle-to-Network-to-Vehicles)[2] communications instead of direct or sidelink V2V (Vehicle-to-Vehicle) communications. These capabilities could be extended in the future with the use of Reconfigurable Intelligent Surfaces (RIS) [1]. Using V2N and V2N2V communications also allows offloading processing tasks from vehicles to the infrastructure [2]. Several studies and trials have started analyzing the latency achieved over the 5G Uu interface. For example, [3] shows that the radio latency of 5G NR (i.e., the latency over the RAN) can be lower than 2 ms in uplink (UL) and downlink (DL) cellular connections. The study also shows that radio latency values lower than 1 ms can be achieved with high SCS and short slot durations. In line with [3], [4] presents a numerical evaluation of different 5G NR configurations and shows that the DL radio latency can be as low as 0.23 ms when 5G NR is configured with a SCS of 60 kHz and mini-slots with only 2 symbols OFDM (Orthogonal Frequency Division Multiplexing), and retransmissions are not considered. The 5G radio latency increases to 3.78 ms when a packet is retransmitted, transmissions use a 14-symbol full-slot, and 5G NR is configured with the lowest SCS (15 kHz). The radio latency values reported in [3] and [4] are obtained when considering small packets and an unlimited bandwidth. Additional evaluations are hence necessary with variable packet sizes and limited bandwidth to consider more realistic scenarios. In this context, [5] showed by simulation that the latency experienced in UL when using dynamic scheduling can increase with the numerology due to signaling exchange necessary for some data traffic patterns. In [6], an existing 5G network from central London is used as a baseline for system-level simulations to evaluate the 5G radio latency. [6] reports radio latencies (UL or DL) of 1 and 5 ms using semi-static scheduling with mini-slots of 4 and 7 OFDM symbols,

UMH work was supported in part by MCIN/AEI/10.13039/501100011033 (grants IJC2018-036862-I, PID2020-115576RB-I00) and by Generalitat Valenciana (GV/2021/044, CIGE/2021/096).

M.C. Lucas-Estañ, B. Coll-Perales, J. Gozalvez and M. Sepulcre are with Universidad Miguel Hernández de Elche, Spain. Contact emails: {m.lucas, bcoll, j.gozalvez, msepulcre}@umh.es. T. Shimizu, T. Higuchi, S. Avedisov, and O. Altintas are with InfoTech Labs, Toyota Motor North America R&D, Mountain View, CA, U.S.A. Contact emails: {takayuki.shimizu, takamasa.higuchi, sergei.avedisov, onur.altintas}@ toyota.com.

[1] Table I presents a list of acronyms used in this paper.
[2] V2N and V2N2V are also referred to as V2C (Vehicle-to-Cloud) and V2C2V to reflect the processing of V2X packets at edge or center cloud.

TABLE I. ACRONYMS.

| Acronym | Meaning | Acronym | Meaning | Acronym | Meaning |
|---------|---------|---------|---------|---------|---------|
| 5G | 5th Generation | LEP | Low error protection | QPSK | Quadrature Phase Shift Keying |
| 5QI | 5G Quality of Service Identifier | LLoA | Low level of automation | RAN | Radio Access Network |
| BLER | Block Error Rate | MCS | Modulation and Coding Scheme | RB | Resource Block |
| CAV | Connected and Automated Vehicles | MIMO | Multiple-Input and Multiple-Output | SCS | Subcarrier spacing |
| CP | Cyclic Prefix | NACK | Negative acknowledgement | SPS | Semi-Persistent Scheduling |
| CQI | Channel Quality Indicator | NCP | Normal Cyclic Prefix | SR | Scheduling request |
| DCI | DL Control Information | NR | New Radio | TBS | Transport Block Size |
| DL | Downlink | OFDM | Orthogonal Frequency Division Multiplexing | TDD | Time Division Duplex |
| ECP | Extended Cyclic Prefix | OS | OFDM symbols | UCI | UL control information |
| FDD | Frequency Division Duplex | PDB | Packet Delay Budget | UE | User Equipment |
| FIFO | First in, first out | PDCCH | Physical DL Control Channel | UL | Uplink |
| FR | Frequency Range | PDSCH | Physical Downlink Shared Channel | V2C | Vehicle-to-Cloud |
| gNB | Next generation Node B | PUCCH | Physical UL Control Channel | V2C2V | Vehicle-to-Network-to-Cloud |
| HARQ | Hybrid-Automatic Repeat reQuest | PUSCH | Physical Uplink Shared Channel | V2N2V | Vehicle-to-Network-to-Vehicles |
| HEP | High error protection | QAM | Quadrature Amplitude Modulation | V2V | Vehicle-to-Vehicle |
| HLoA | High level of automation | QoS | Quality of Service | V2X | Vehicle-to-Everything |

respectively, and a SCS of 30 kHz for short packets of 32 bytes. [7] evaluates the 5G UL radio latency using dynamic and semi-static scheduling and shows that values lower than 1 ms can only be achieved with an SCS of 15 kHz when using mini-slots transmissions with only 4 or 2 OFDM symbols and without retransmissions. The study shows that radio latencies below 1 ms are also possible with (mini-/full-) slots with 7 and 14 OFDM symbols, one retransmission and semi-static scheduling when using higher SCS. The study shows that the latency introduced by the signaling present in dynamic scheduling result in latency values above 1 ms if retransmissions are considered. The field trials reported in [8] also show UL+DL radio latency values below 2 ms using 60 kHz SCS for connecting 3 trucks in a platoon. We should note that existing studies generally do not address the scalability of 5G NR. Such studies have been reported for LTE in [9] where authors analyzed how the latency of LTE is affected when the network scales and quantified its degradation with the network load. However, it is still necessary to understand whether 5G can provide the low latency and high reliability performance that certain V2X services require as the traffic load on the network increases. The authors presented in [10] and [11] first studies that show how the V2X traffic load can significantly impact the 5G radio latency as the network scales and the allocated spectrum is not unlimited. [10] and [11] were constrained to a specific 5G NR configuration.

Existing studies provide valuable information on the 5G radio latency that can be achieved over the 5G NR Uu interface. However, there are still multiple 5G NR features and configurations that need to be analyzed to assess the potential of 5G to support advanced V2X services and identify which 5G NR configurations better support these services. For example, it is necessary to evaluate the use of semi-static and dynamic scheduling for different traffic types (periodic or aperiodic) and under different traffic loads. 5G NR can increase reliability using HARQ retransmissions or transmitting multiple replicas of each packet, but these techniques are not exempt of tradeoffs (e.g., latency or spectrum efficiency) that should be analyzed. To this aim, this paper advances the state-of-the art by presenting a versatile analytical model that can be used to quantify the 5G NR radio latency over the RAN for a wide set of parameters and the most relevant 5G NR configurations and features for supporting advanced V2X services. This is in contrast to previous analytical studies ([3], [4] and [7]) that generally focused on specific 5G NR configurations, including unlimited bandwidth and the transmission of small packets. Our model accounts for different SCS, slot durations, modulation and coding schemes, scheduling, and retransmission schemes in 5G NR. The model also considers the impact of the allocated bandwidth, the characteristics of the data traffic, and the density and distribution of vehicles in the cell on the 5G NR radio latency. Our model is therefore suitable to evaluate the scalability of the 5G NR radio network. The proposed model is used in this study to evaluate the radio latency and reliability that can be achieved with different 5G NR configurations and schemes when supporting connected and automated driving V2X services with diverse latency and reliability requirements based on their level of automation. The implementation of the proposed latency models is openly available at [12].

## II. 5G NEW RADIO

5G NR uses OFDM and defines various numerologies that are identified by a different SCS in the frequency domain and slot duration in the time domain. Numerologies $\mu$ equal to 0, 1 and 2 using 15, 30, and 60 kHz SCS are used for the lower frequency range (FR1, 410 MHz-7.125 GHz), and $\mu$ equal to 2, 3 and 4 using 60, 120, and 240 kHz SCS are used for the higher frequency range (FR2, 24.25 GHz-52.6 GHz). 12 consecutive subcarriers form a Resource Block (RB) in the frequency domain. In the time domain, the duration of a slot is equal to $1/2^{\mu}$ ms. 5G NR supports a normal cyclic prefix (NCP) for all numerologies, and an extended cyclic prefix (ECP) for numerology 2. A slot consists of 14 or 12 OFDM symbols when NCP or ECP are used, respectively. DL and UL transmissions are organized into frames with a duration of 10 ms.

5G NR supports Time Division Duplex (TDD) and Frequency Division Duplex (FDD). TDD offers higher flexibility as each slot within the frame can be configured for DL or UL transmissions. On the other hand, FDD is implemented on a paired spectrum with DL and UL transmissions sent on separate frequencies. This can reduce the latency as there are resources available simultaneously for DL and UL transmissions with FDD. 5G NR offers the possibility to transmit using full-slots (i.e., 14 or 12 OFDM symbols) or mini-slots (1 to 13 symbols in UL, and 2 to 13 symbols in DL).

5G NR uses Low Density Parity Check coding schemes for data channels together with Quadrature Phase Shift Keying (QPSK), 16 Quadrature Amplitude Modulation (QAM), 64QAM or 256 QAM modulations. 5G NR defines three different MCS (Modulation and Coding Scheme) index tables in [13] for data transmissions in the DL using the Physical Downlink Shared Channel (PDSCH). MCS Tables 1 and 3 in [13] consider only the use of QPSK, 16QAM and 64QAM modulations, while Table 2 also considers 256QAM. MCS Tables 1 and 2 in [13] have been defined to provide high spectrum efficiency (up to 5.5547 and 7.4063 bit/s/Hz respectively), and can achieve a target Block Error Rate (BLER) of 0.1 when used following the CQI (Channel Quality Indicator) Tables 1 and 2 defined in [13], respectively. The CQI Tables identify the maximum MCS that can achieve a target BLER (0.1 in this case) for a reported CQI value. MCS Table 3 uses lower coding rates and more robust MCSs at the expense of a lower spectrum efficiency of 4.5234 bit/s/Hz and can achieve a target BLER of $10^{-5}$ or lower if used according to the CQI Table 3 defined in [13]. The same three MCS index tables are also used for UL data transmissions.

5G NR introduces dynamic and semi-static scheduling for both DL and UL transmissions [14]. The dynamic scheme allocates new resources for each transmission. The gNB informs the UE whenever a packet for DL is generated and the radio resources to use to receive the packet. When a packet is generated for UL, the UE sends a scheduling request to the gNB, and the gNB replies with an UL grant and the radio resources to use to transmit the packet. Dynamic scheduling guarantees an efficient management of the radio resources that is particularly useful for aperiodic traffic. However, it can increase the latency due to the signaling exchange between the UE and the gNB that can be detrimental to certain safety-critical V2X services. 5G NR also offers the possibility to use semi-static scheduling schemes [14]: Semi-Persistent Scheduling (SPS) for DL and Configured Grant for UL. These schemes pre-assign resources periodically for data transmissions before packets are generated. The gNB informs the UEs about the pre-allocated resources and their periodicity. Then, UEs do not need to request resources each time they have a new packet to transmit, and they know in advance which RBs they can utilize for their transmission; this could ultimately improve the latency. 3GPP defines in [14] two types of Configured Grant: type 1 where the configured uplink grant is active since the moment it is configured, and type 2 where the configured uplink grant can be activated or deactivated (type 2 is like SPS in DL). Semi-static scheduling schemes can reduce the transmission latency but can also entail certain inefficiencies in the utilization of the spectrum if pre-assigned resources are not used by a UE (e.g., with aperiodic traffic).

5G NR includes two retransmission mechanisms to increase reliability. The first one is asynchronous Hybrid Automatic Repeat reQuest (HARQ) retransmissions [14]. With HARQ, a packet is retransmitted if the receiver negatively acknowledges the reception of a packet. HARQ retransmissions improve the reliability at the cost of latency due to the exchange of messages between the transmitter and receiver. The latency can be reduced with the transmission of $k$ replicas or repetitions of the same packet in consecutive slots ($k$ can be equal to 2, 4 or 8) [15]. This second approach does not require transmitters and receivers to exchange acknowledgement messages. However, the use of $k$-repetitions can reduce spectrum efficiency since some of these repetitions might be unnecessary if a previous transmitted repetition of a packet was successfully received.

We should note that 5G Release 16 does not yet support broadcast and multicast communications, which may be critical for the scalability of certain V2X services. However, 3GPP is studying and evaluating under Release 17 enhancements to the 5G system architecture [16] and to the 5G NR RAN [17] to support both communication modes.

## III. 5G Radio Latency Models

This section presents the analytical model derived to estimate the latency of 5G NR communications at the RAN (or radio latency). In particular, we estimate the radio latency $l_{radio}$ over the Uu interface for UL and DL communications. This study focuses on V2N2V communications, so we define $l_{radio}$ as the sum of the latency $l_{radio}^{UL}$ experienced from the UE to the gNB, and the latency $l_{radio}^{DL}$ experienced from the gNB to the UE. We estimate the radio latency for different 5G NR configurations (numerology, slot duration, MCSs, scheduling, retransmission schemes and transmission mode), and considering the allocated bandwidth, the characteristics of the data traffic, and the density and distribution of vehicles in the cell.

$l_{radio}^x$ (with $x$ corresponding to UL or DL) is estimated considering the latency $T_{ini}^x$ resulting from the initial transmission of a packet, and the latency $n \cdot T_r^x$ resulting from the transmission of the $n$ additional repetitions of the packet or the $n$ potential HARQ retransmissions when applicable. We can then express $l_{radio}^x$ as:

$$l_{radio}^x = T_{ini}^x + n \cdot T_r^x \text{ with } x = \text{UL, DL} \qquad (1)$$

$T_{ini}^x$ depends on the latency introduced by the scheduling process $t_{sch}^x$, that represents the time between the time instant a UE or gNB has data to transmit and the time instant the UE is notified by the gNB of the radio resources to use for the packet transmission, and the latency $t_{pkt}^x$ that represents the time from when the UE receives the information about the resources to use to the time instant when the packet is received at the destination. $T_{ini}^x$ is then expressed as:

$$T_{ini}^x = t_{sch}^x + t_{pkt}^x \qquad (2)$$

$T_r^x$ in (1) depends on whether HARQ or $k$-repetitions are used. It can be expressed as the sum of the latency $t_{r-req}^x$ resulting from the signaling exchanged to request a new transmission of a packet, and the latency $t_{r-rep}^x$ corresponding to the new transmission of the packet:

$$T_r^x = t_{r-req}^x + t_{r-rep}^x. \qquad (3)$$

Like $T_{ini}^x$, $t_{r-rep}^x$ also depends on the scheduling scheme.

The model analyzes $l_{radio}$ for transmissions using any combination of scheduling (semi-static or dynamic), retransmission mechanism (HARQ or $k$-repetitions), and transmission mode (unicast or broadcast/multicast). First, we model the radio latency experienced by a packet when it is transmitted only once (i.e., no retransmissions or repetitions are considered) using semi-static scheduling (Configured Grant in UL and SPS in DL). We also model the case when each packet can only be transmitted once but using dynamic scheduling in both UL and DL. Then, we analyze the latency experienced by a packet that is scheduled using semi-static or dynamic scheduling, when $k$ repetitions of the packet are transmitted in consecutive slots ($k$-repetitions). We also model the latency experienced by a packet that is scheduled using semi-static or dynamic scheduling, when the packet can be retransmitted using HARQ. All these combinations of scheduling and retransmissions schemes consider that for each packet transmitted in UL, one packet is broadcast or multicast in DL so that multiple vehicles can receive it. We considered broadcast/multicast communications (even if not available under Release 16) since they will be essential for supporting many V2X services using V2N or V2N2V communications. We also derive an analytical model of the radio latency when using unicast communications for the DL in line with Release 16. In this case, the gNB sends multiple unicast transmissions in DL to reach neighboring vehicles for each packet received in UL. Unicast transmissions can be performed using semi-static or dynamic scheduling, with $k$-repetitions or HARQ retransmissions. We should note that models for all these combinations are derived considering that packets can be transmitted using full-slots or mini-slots with different number of symbols. The models have been derived for FDD since FDD can reduce the latency compared to TDD as radio resources are simultaneously available for UL and DL. We should note that 5G NR FDD frequency bands have been assigned or are planned worldwide.

### A. Single packet transmission using Configured Grant in UL and SPS in DL

First, we consider that packets are only transmitted once and there are no retransmissions or repetitions, i.e. $n \cdot T_r^x = 0$ in (1). Packets are transmitted in the UL and DL using Configured Grant and SPS, respectively. Both scheduling schemes preallocate radio resources periodically for each UE when the session is established. The DL resources are used to broadcast/multicast the UL packets to neighboring vehicles. Since resources are pre-allocated, the latencies introduced by the scheduling in UL ($t_{sch-CG}^{UL}$) and DL ($t_{sch-SPS}^{DL}$) are both equal to zero. Following (1) and (2), $l_{radio}^x$ is then equal to $t_{pkt}^x$. To compute $t_{pkt}^x$, we should take into account the latency components represented in Fig. 1 for the DL (the same applies to the UL): 1) the processing delays at the transmitter and the receiver ($t_p^{tx-UE}$, $t_p^{tx-gNB}$, $t_p^{rx-UE}$ and $t_p^{rx-gNB}$ when the UE and the gNB are the transmitters or the receivers, respectively); 2) the frame alignment times ($t_{fa}$); 3) the waiting time for the allocated resources or RBs ($t_w$); and 4) the transmission time ($t_{tt}$). $l_{radio}^{UL}$ and $l_{radio}^{DL}$ can then be expressed as:

$$l_{radio}^{UL} = T_{ini}^{UL} = t_{pkt}^{UL} = t_p^{tx-UE} + t_{fa} + t_w + t_{tt} + t_p^{rx-gNB} \tag{4}$$

$$l_{radio}^{DL} = T_{ini}^{DL} = t_{pkt}^{DL} = t_p^{tx-gNB} + t_{fa} + t_w + t_{tt} + t_p^{rx-UE} \tag{5}$$

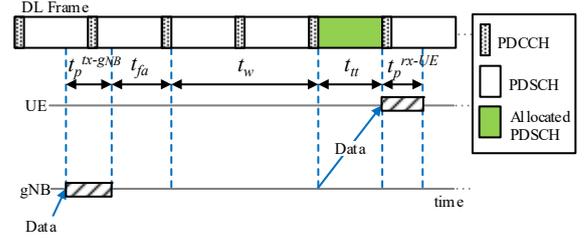

Fig. 1. $t_{pkt}^{DL}$ experienced in the DL when using SPS (striped rectangles represent the processing of packets in the UE or gNB).

At the transmitter, the processing delay represents the time interval between the generation of the data and the creation of the packet. At the receiver, the processing delay represents the time interval between the reception of a packet and the decoding of the data in the packet. Following [3], $t_p^{tx-UE} = t_p^{tx-gNB} = T_{proc,2}/2$ and $t_p^{rx-UE} = t_p^{rx-gNB} = T_{proc,1}/2$. $T_{proc,1}$ represents the UE PDSCH processing procedure time, and $T_{proc,2}$ represents the PUSCH preparation procedure time. $T_{proc,1}$ and $T_{proc,2}$ depend on the numerology and the UE processing capabilities [13].

The term $t_{fa}$ accounts for the time interval from the creation of a packet until the start of the next transmission opportunity for the PUSCH or PDSCH (see Fig. 1). $t_{fa}$ depends on the frame structure. We compute the exact value of $t_{fa}$ for each packet by emulating the generation of data packets at the transmitter. Following [18], the Physical DL Control Channel (PDCCH) is always transmitted in the first symbols of each slot in the DL FDD frame, and the rest of the symbols of each slot are used for the transmission of the PDSCH. In the UL, the Physical UL Control Channel (PUCCH) is always transmitted in the last symbols of each slot, and the rest of the symbols are used for the transmission of the PUSCH. Therefore, $t_{fa}$ is bounded by the slot duration, i.e., 1 ms to 0.0625 ms depending on the numerology.

After $t_{fa}$, $t_w$ accounts for the waiting time until the RBs allocated to the UE for data transmission become available (Fig. 1). To estimate $t_w$, we emulate the resource allocation process. To this end, we need to know: 1) the number of available RBs and 2) the number of RBs and symbols $N_{sy}$ needed to transmit the packet. The number of available RBs depends on the bandwidth and the numerology and is specified in 3GPP TS 38.104. The number of symbols $N_{sy}$ needed to transmit a packet depends on whether full-slots or mini-slots are used. The number of RBs necessary to transmit a packet is given by the transport block size (TBS) that is calculated following the procedure defined in [13]. TBS depends on the size of the data to transmit, the MCS and the number of Multiple-Input and Multiple-Output (MIMO) transmission layers or simultaneous data streams used to transmit the packet. We consider the use of the MCSs defined in the MCS Tables 1, 2 and 3 in [13] for PDSCH and PUSCH. The MCS is adapted as a function of the estimated CQI using the CQI Tables 1, 2 or 3 in [13]. Once we know the number of RBs and symbols utilized to transmit the packet, we emulate the resource allocation process to identify the available free RBs per symbol and allocate the RBs for the transmission of the packet (we exclude out of the total number of RBs those used for control channels and physical layer signals). We can then estimate $t_w$ considering the number of vehicles in the cell and the traffic model. We consider that the scheduler allocates RBs in the first slot where it can find the necessary consecutive free symbols and RBs to transmit the packet. Note that a mini-slot transmission can start at any symbol of a slot and can only use symbols of this slot. The transmission time $t_{tt}$ is equal to the time duration of the symbols used to transmit the packet (Fig. 1). $t_{tt}$ depends then on the numerology and $N_{sy}$.

### B. Single packet transmission using dynamic scheduling

We now consider that each packet is transmitted only once (i.e., $n \cdot T_r^x = 0$ in (1)), but the radio resources are allocated using dynamic scheduling. $l_{radio}^{UL}$ and $l_{radio}^{DL}$ (Fig. 2) are computed as the sum of the latency introduced by dynamic scheduling ($t_{sch-DS}^{UL}$ in UL and $t_{sch-DS}^{DL}$ in DL), and $t_{pkt}^{UL}$ and $t_{pkt}^{DL}$ (estimated in (4) and (5)):

$$l_{radio}^{UL} = T_{ini}^{UL} = t_{sch-DS}^{UL} + t_{pkt}^{UL} \qquad (6)$$

$$l_{radio}^{DL} = T_{ini}^{DL} = t_{sch-DS}^{DL} + t_{pkt}^{DL} \qquad (7)$$

Dynamic scheduling in DL requires the gNB to inform the UE when it has data to receive and provides the UE with the necessary scheduling information (frequency and time resources assigned and MCS to use, among other parameters) over the PDCCH (Fig. 2.a). The latency introduced by the scheduling $t_{sch-DS}^{DL}$ includes: 1) the processing delay $t_p^{tx-PDCCH}$ in the gNB; 2) the frame alignment time $t_{fa}$ for transmission of the PDCCH; 3) the queuing delay $t_q^{PDCCH}$; 4) the transmission time $t_{tt}^{PDCCH}$ of the PDCCH that is equal to the duration of the symbols used to transmit the PDCCH; and 5) the processing delay $t_p^{rx-PDCCH}$ in the UE. $t_{sch-DS}^{DL}$ can then be expressed as:

$$t_{sch-DS}^{DL} = t_p^{tx-PDCCH} + t_{fa} + t_q^{PDCCH} + t_{tt}^{PDCCH} + t_p^{rx-PDCCH} \qquad (8)$$

The terms $t_p^{tx-PDCCH}$ and $t_p^{rx-PDCCH}$ represent the time needed to generate and decode the PDCCH, respectively. Following [3], $t_p^{tx-PDCCH} = T_{proc,1}/2$ and $t_p^{rx-PDCCH} = T_{proc,2}/2$. To estimate the queuing delay $t_q^{PDCCH}$, we emulate the generation and queuing processes of the DL Control Information (DCI) messages considering the number of vehicles (or UEs) $N_{UE}$ in the cell and the average rate $\lambda_{pkt}$ at which each UE generates data packets. DCI messages contain the DL scheduling information and are transmitted by the gNB on the PDCCH in FIFO (first in, first out) order. When the DCI messages cannot be transmitted on the next PDCCH, they are queued at the gNB waiting for the next PDCCH transmission opportunity. $t_q^{PDCCH}$ then accounts for the time the message is queued at the gNB until it is transmitted. The number of DCI messages that can be transmitted on each PDCCH depends on the number of resources reserved for the PDCCH and the size of the DCI. We consider that there are $N_{RB}^{PDCCH}$ RBs reserved in the first $N_{sy}^{PDCCH}$ symbols of each slot for the PDCCH, and that the DCI message with the DL scheduling information is transmitted using DCI format 1_0. The size of the DCI message using DCI format 1_0 is equal to or lower than 68 bits (3GPP TS 38.212 V16.1.0) and can be transmitted using 6 RBs in 1 symbol. The number of DCI

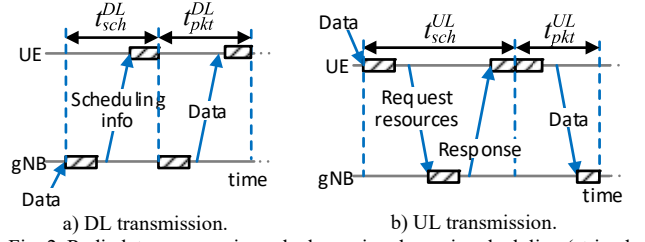

a) DL transmission.   b) UL transmission.
Fig. 2. Radio latency experienced when using dynamic scheduling (striped rectangles represent the processing of packets in the UE or gNB).

messages that can be transmitted at each PDCCH is then equal to $N_{RB}^{PDCCH} \cdot N_{sy}^{PDCCH}/6$. To estimate $t_q^{PDCCH}$, we also need to compute the rate at which DCI messages are generated in the cell. A DCI message is generated per data packet, so the DCI message generation rate can be estimated as $N_{UE} \cdot \lambda_{pkt}$.

In the UL, the UE must send a scheduling request (SR) to the gNB when it has data to transmit, and the gNB replies with a message that contains the grant with the UL scheduling information. $t_{sch,DS}^{UL}$ can then be expressed as:

$$t_{sch,DS}^{UL} = t_{SR} + t_{grant} \qquad (9)$$

where $t_{SR}$ and $t_{grant}$ represent the latency introduced by the transmission of the scheduling request and grant, respectively. The scheduling request is transmitted on the PUCCH, and $t_{SR}$ accounts for (Fig. 2.b): 1) the PUCCH processing delay $t_p^{tx-PUCCH}$ at the UE; 2) the frame alignment time $t_{fa}$; 3) the waiting time $t_w^{PUCCH}$ for the PUCCH resources allocated to the UE to transmit the SR; 4) the transmission time of the PUCCH $t_{tt}^{PUCCH}$; and 5) the PUCCH processing delay $t_p^{rx-PUCCH}$ at the gNB. $t_{SR}$ is then calculated as:

$$t_{SR} = t_p^{tx-PUCCH} + t_{fa} + t_w^{PUCCH} + t_{tt}^{PUCCH} + t_p^{rx-PUCCH} \qquad (10)$$

where $t_p^{tx-PUCCH} = T_{proc,1}/2$ and $t_p^{rx-PUCCH} = T_{proc,2}/2$ [13].

To estimate $t_w^{PUCCH}$, we should note that the gNB informs each UE about the PUCCH resources it can access to transmit its SRs that are contained in UL control information (UCI) messages. $t_w^{PUCCH}$ depends on: 1) the number of resources reserved for the PUCCH; 2) the number of resources needed to transmit a UCI message with the SR; and 3) the number $N_{UE}$ of UEs in the cell. We consider that there are $N_{RB}^{PUCCH}$ RBs by $N_{sy}^{PUCCH}$ symbols reserved for the PUCCH at the end of each slot, and that the SR is transmitted using one RB in one symbol (SR uses PUCCH format 0 that can multiplex up to 6 UEs 3GPP TS 38.300). The number $R_{SR}$ of SR messages that can be transmitted per slot is then equal to $R_{SR} = N_{RB}^{PUCCH} \cdot N_{sy}^{PUCCH} \cdot 6$. A UE is then allocated resources for the transmission of the PUCCH every $N_{slots}^{SR} = \lceil N_{UE}/R_{SR} \rceil$ slots (where $\lceil \cdot \rceil$ is the ceiling function), and the probability that a UE can transmit the SR in a slot is given by $1/N_{slots}^{SR}$. $t_w^{PUCCH}$ can then be estimated as:

$$t_w^{PUCCH} = \begin{cases} 0 & \text{if } p = 0 \\ T_{slot} \cdot ([p \cdot N_{slots}^{SR}] - 1) & \text{if } p > 0 \end{cases} \qquad (11)$$

where $p$ is a random number between 0 and 1.

The UL grant sent by the gNB to the UE is contained in the DCI transmitted on the PDCCH. $t_{grant}$ is estimated using (8).

### C. Transmission of k repetitions of a packet

We focus now on estimating the latency experienced in the transmission of a data packet when $k$ repetitions are transmitted in consecutive slots. Any of the scheduling schemes analyzed in the two previous configurations can be used to transmit the $k$ repetitions of a packet. The latency experienced in the initial transmission of the packet $T_{ini}^{UL}$ in UL and $T_{ini}^{DL}$ in DL ($T_{ini}^x$ in (1)) is equal to $l_{radio}^{UL}$ in (6) and $l_{radio}^{DL}$ in (7), respectively, when using dynamic scheduling. $T_{ini}^{UL}$ and $T_{ini}^{DL}$ are equal to $l_{radio}^{UL}$ in (4) and $l_{radio}^{DL}$ in (5) when using Configured Grant in UL and SPS in DL. The transmitter does not wait for any request from the receiver to transmit the following $k$-1 repetitions of the packet. $t_{r-req}^{UL}$ and $t_{r-req}^{DL}$ in (3) are then equal to 0. $t_{r-rep}^{UL}$ and $t_{r-rep}^{DL}$ in (3) are equal to $T_{slot}$ since each repetition is transmitted in consecutive slots. $l_{radio}^{UL}$ and $l_{radio}^{DL}$ can then be expressed as:

$$l_{radio}^{UL} = T_{ini}^{UL} + (k\text{-}1) \cdot T_{slot} \tag{12}$$

$$l_{radio}^{DL} = T_{ini}^{DL} + (k\text{-}1) \cdot T_{slot} \tag{13}$$

### D. Use of HARQ retransmissions

We now consider the case in which each packet can be retransmitted up to $n$ times using HARQ retransmissions. HARQ can be applied in DL unicast or multicast; 3GPP already agreed in Rel-17 to support HARQ retransmissions for multicast transmissions. Any of the scheduling schemes analyzed in the first two cases can be used to transmit the initial packet. Dynamic scheduling is used for the retransmissions since it is not possible to anticipate if (and when) the receiver(s) might request a retransmission. The latency experienced by the initial transmission of a packet $T_{ini}^{UL}$ in UL and $T_{ini}^{DL}$ in DL defined in (1) is equal to $l_{radio}^{UL}$ in (4) or $l_{radio}^{DL}$ in (5) when using Configured Grant in UL and SPS in DL, and equal to $l_{radio}^{UL}$ in (6) or $l_{radio}^{DL}$ in (7) when using dynamic scheduling.

A packet is retransmitted if it is not successfully received, and the receiver sends a negative acknowledgement or NACK. The transmission of the NACK introduces a latency ($t_{NACK}^{UL}$ or $t_{NACK}^{DL}$). The NACK is transmitted on the PUCCH in UL and on the PDCCH in DL. $t_{NACK}^{UL}$ and $t_{NACK}^{DL}$ must then consider: 1) the processing delay $t_p^{tx\text{-}PUCCH}$ or $t_p^{tx\text{-}PDCCH}$; 2) the frame alignment time $t_{fa}$; 3) the transmission time of the PUCCH ($t_{tt}^{PUCCH}$) or PDCCH ($t_{tt}^{PDCCH}$); and 4) the processing time ($t_p^{rx\text{-}PUCCH}$ or $t_p^{rx\text{-}PDCCH}$) needed to decode the PUCCH or the PDCCH. $t_{NACK}^{UL}$ and $t_{NACK}^{DL}$ can then be estimated as:

$$t_{NACK}^{UL} = t_p^{tx\text{-}PUCCH} + t_{fa} + t_{tt}^{PUCCH} + t_p^{rx\text{-}PUCCH} \tag{14}$$

$$t_{NACK}^{DL} = t_p^{tx\text{-}PDCCH} + t_{fa} + t_{tt}^{PDCCH} + t_p^{rx\text{-}PDCCH} \tag{15}$$

The retransmission of a packet (i.e. $t_{r-rep}^x$ in (3)) using dynamic scheduling introduces a latency that is given by $t_{sch,DS}^{UL} + t_{pkt}^{UL}$ ((9) and (4)) in UL and by $t_{sch,DS}^{DL} + t_{pkt}^{DL}$ ((8) and (5)) in DL. We can then express $l_{radio}^{UL}$ and $l_{radio}^{DL}$ as:

$$l_{radio}^{UL} = T_{ini}^{UL} + n \cdot T_{r\text{-}HARQ}^{UL} = T_{ini}^{UL} + n \cdot (t_{NACK}^{UL} + t_{sch,DS}^{UL} + t_{pkt}^{UL}) \tag{16}$$

$$l_{radio}^{DL} = T_{ini}^{DL} + n \cdot T_{r\text{-}HARQ}^{DL} = T_{ini}^{DL} + n \cdot (t_{NACK}^{DL} + t_{sch,DS}^{DL} + t_{pkt}^{DL}) \tag{17}$$

### E. Multiple DL unicast transmissions

The latency models presented in the previous subsections have been derived considering broadcast or multicast transmissions in DL. Since 3GPP Release 16 does not yet include broadcast and multicast transmissions, we now consider the case in which for each packet that a vehicle transmits in the UL, we perform $M$ unicast transmissions in the DL to distribute the data among $M$ neighboring vehicles. We consider that the same radio configuration (SCS, cyclic prefix or CP, MCS, scheduling and retransmission schemes) is used for the $M$ DL transmissions. However, all $M$ DL transmissions are independent, and their scheduling and potential retransmissions are individually managed. $l_{radio}^{UL}$ can be computed using (4), (6), (12), or (16) depending on the scheduling and retransmission schemes utilized following Sections III.A to III.D. The latency $l_{radio,m}^{DL}$ (with $m$=1,...,$M$) experienced by each of the $M$ DL unicast transmissions is also calculated using (5), (7), (13), or (17). We then compute the latency $l_{radio}^{DL}$ as equal to the largest value from the set of $l_{radio,m}^{DL}$ values, with $m$=1,...,$M$:

$$l_{radio}^{DL} = \max \{l_{radio,1}^{DL}, ..., l_{radio,M}^{DL}\} \tag{18}$$

## IV. EVALUATION SCENARIO

This section presents the scenario considered to evaluate the capabilities of 5G NR at the RAN to support V2X services using V2N2V communications. We consider a 5G NR cell as in the reference 3GPP study [3]. The cell operates in FDD mode and is assigned a bandwidth of $BW$ for UL and DL respectively in Frequency Range 1; unless stated otherwise, $BW$ is set equal to 20 MHz. An SCS of 30 kHz SCS is assumed in [19] for the transportation industry. However, this study also evaluates the performance with 15 and 60 kHz SCS. ECP is used with 60 kHz SCS for better combating Inter-symbol Interference, while NCP is used with 15 and 30 kHz SCSs. We consider full-slot as well as mini-slot (with 4 or 7 symbols) transmissions. We consider UEs with processing capability 2 since it has shorter PDSCH and PUSCH processing times [13]. Based on [19], UEs are equipped with 2 transmitting and receiving antennas, and the number of transmission layers is limited to 2. UL and DL transmissions use the MCSs defined in MCS Tables 2 and 3 in [13], and the MCS is adapted as a function of the CQI to achieve a target BLER of 0.1 or $10^{-5}$ (using CQI Tables 2 or 3 in [13] respectively). The CQI is estimated as a function of the distance between the vehicle and the serving gNB. We refer to the MCSs defined in Table 2 as low error protection (LEP) MCSs, and those in Table 3 as high error protection (HEP) MCSs. The number of RBs available in the cell depends on the bandwidth and the numerology as defined in 3GPP TS 38.104. Some of these RBs are used to transmit the control channels and physical layer signals defined in 5G NR. We select the number of RBs reserved for control messages following the set-up in Annex A of [20].

The 5G NR cell has a radius of 866 m and provides coverage to a highway scenario with 3 lanes per direction [21]. Vehicles are distributed following a uniform random distribution. We evaluate the performance under free-flow traffic conditions (vehicle densities equal to 10-20 veh/km/lane), medium traffic densities (40-60 veh/km/lane), and traffic congestion (80 veh/km/lane). The values for the different evaluation parameters are summarized in Table II. Without loss of generality, this study considers the communication requirements identified by 3GPP in [22] for the cooperative lane change V2X service with different levels of automation. The 3GPP makes a distinction in [23] between low level of automation (LLoA) when the driver is primarily responsible for monitoring the driving environment (i.e. SAE automation levels 0-2 [24]), and high level of automation (HLoA) when the automated system is responsible for monitoring the driving environment (i.e. SAE levels 3-5). For LLoA, 3GPP establishes that messages must be exchanged between vehicles with a maximum end-to-end latency of 25 ms and a reliability of 90%. The reliability is defined as the percentage of packets that must satisfy the latency requirement. It is important to highlight that the end-to-end latency considers the latency experienced in the radio, transport, and core networks as well as at the application server, but this study focuses on the latency only at the radio level (interested readers are referred to [25] for the evaluation of the complete end-to-end latency). To establish the latency requirement for the radio network, we consider the 5G Quality of Service Identifier (5QI) 83 defined by the 3GPP in [26]. 3GPP has defined a series of 5QIs to support different services in various verticals. A 5QI is a set of 5G Quality of Service (QoS) parameters that control the QoS forwarding treatment [26]. One of these QoS parameters is the packet delay budget (PDB) that defines an upper bound for the maximum one-way latency that a packet may experience between the UE and the last node in the core network. 5QI 83 supports, among others, the cooperative lane change service with LLoA. [26] establishes that 1 ms is reserved in 5QI 83 for the transmission of a packet through the core network (one-way path). Since the end-to-end latency requirement for the service with LLoA is 25 ms, we consider that the UL+DL radio latency requirement $l_{radio}$ for this service is equal to 23 ms. For the cooperative lane change service with HLoA, we consider 5QI 86 in [26] since it supports advanced driving services (including cooperative lane change). 5QI 86 establishes a PDB of 5 ms, with 2 ms reserved for the core network. 3GPP establishes an end-to-end latency requirement of 10 ms with a reliability of 99.99% for the cooperative lane change with HLoA. We then consider that the UL+DL radio latency ($l_{radio}$) requirement for HLoA is equal to 6 ms. HLoA has more stringent radio latency and reliability requirements than LLoA. We then consider the use of HEP MCSs for the HLoA service, and the use of LEP MCSs for the LLoA service for a lower usage of RBs.

All vehicles supporting the cooperative lane change service exchange packets of 300 bytes [27] with information about the neighboring vehicles' trajectories. Packet generation is emulated for periodic and aperiodic traffic. For periodic traffic, we consider transmission periods $T_p$ of 20 or 100 ms [21]. For aperiodic traffic, the time between packets $\Delta T_a$ is modelled following [21] using an exponential function ($rand\_exp$) with average $T_{avg}/2$, i.e. $\Delta T_a = T_{avg}/2 + rand\_exp(T_{avg}/2)$, where $T_{avg}$ is equal to 20 or 100 ms. It is important that vehicles receive the most updated information about the neighboring vehicles' trajectories. As a result, a transmitter drops a packet if it has not been transmitted by the time the next packet is generated.

TABLE II. EVALUATION PARAMETERS.

| Parameter | Value |
|---|---|
| Scenario | Highway with 3 lanes per direction |
| Vehicle density | 10, 20, 40, 60, 80 veh/km/lane |
| Traffic type | Periodic and aperiodic |
| Avg. time between packets ($T_p$ for periodic traffic and $T_{avg}$ for aperiodic traffic) | 20 and 100 ms |
| Packet size | 300 bytes |
| Cell radius | 866 m |
| Duplex mode | FDD |
| Frequency Range | FR1 |
| UL and DL Bandwidth ($BW$) | 10, 20, 30, 40 and 50 MHz |
| SCS and CP | 15 and 30 kHz NCP, 60 kHz ECP |
| Transmission slot | Full-slot and mini-slot (4 and 7 OFDM symbols) |
| UE processing capability | 2 |
| Transmission layers | 2 |

## V. EVALUATION

This section evaluates the radio latency achieved with different 5G NR configurations when supporting the cooperative lane change service with LLoA and HLoA using V2N2V communications over the 5G network. The section analyzes the impact of these configurations on the capacity to satisfy the service requirements when the service generates periodic or aperiodic traffic. The V2N2V radio latency is evaluated using the models previously derived[3].

### A. Periodic traffic

We first evaluate the V2N2V radio latency when the cooperative lane change service generates packets periodically. In this case, UL and DL packets are scheduled with Configured Grant and SPS respectively for an efficient use of the RBs.

#### 1) Unicast DL transmissions

We first analyze the V2N2V radio latency when DL transmissions are executed using multiple unicast transmissions to reach neighboring vehicles (Section III.E). We also consider first that packets are only transmitted once and there are no retransmissions. Fig. 3 depicts the average V2N2V radio latency $\overline{l_{radio}}$ as a function of the traffic density. A transmitter drops a packet if it has not been transmitted when a new one is generated. $\overline{l_{radio}}$ accounts for the latency of the packets that are transmitted and not dropped. $M$ is the number of neighboring vehicles addressed in the DL after a vehicle has transmitted a packet in the UL. Results are shown for $T_p$=20 ms and $T_p$=100 ms, and for cooperative lane change with LLoA and HLoA. LLoA uses LEP MCSs whereas HLoA uses HEP MCSs.

Fig. 3 shows that higher V2N2V radio latency $\overline{l_{radio}}$ values are observed for the HLoA service than for the LLoA one. This is due to the use of HEP MCSs that increase the error protection, and hence require a higher number of RBs to transmit a data packet than is required for LEP MCSs. On average, 6.6 RBs are needed to transmit a 300-byte packet with HEP MCSs, while only 2.5 RBs are needed with LEP MCSs. Augmenting the number of RBs per packet with the HEP MCSs increases the probability of congesting the cell, and hence the percentage of dropped packets. For example, the cell is congested (i.e. more than 99% of the RBs are used) for $M$=4 and $M$=6 when considering the HLoA service with HEP MCSs for densities equal to or higher than 40 veh/km/lane and $T_p$=100 ms. On the other hand, only 42% and 62% of RBs are used for $M$=4 and $M$=6, respectively, with LLoA (and LEP MCSs) when the density is equal to 40 veh/km/lane and $T_p$=100 ms. These differences explain why the HLoA service experiences a significantly higher percentage of dropped packets with the traffic density compared to LLoA.

---

[3] The latency components $t_{fa}$, $t_w$, and $t_q^{PDCCH}$ are estimated by means of simulations in Matlab since these three components depend on the generation of packets, the resource allocation process, and the queuing of DCI messages respectively. The source code is openly available at [12]. We run sufficient simulations to achieve statistically valid results (i.e. results with a relative error below 1% for each evaluated configuration).

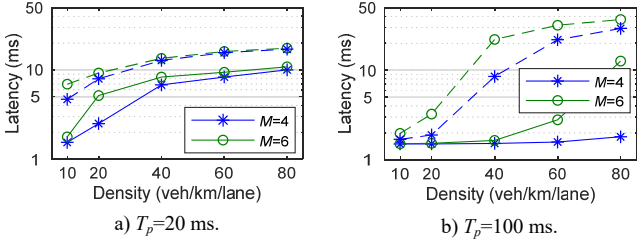

Fig. 3. V2N2V $\overline{l_{radio}}$ as a function of the density with unicast DL, LLoA (solid lines) and HLoA (dashed lines) (full-slot, SCS=30 kHz, $BW$=20 MHz).

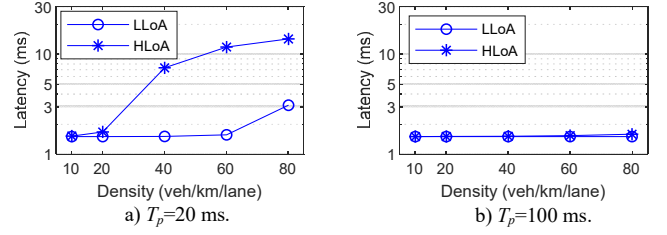

Fig. 4. Average V2N2V $\overline{l_{radio}}$ as a function of the traffic density with broadcast DL transmissions (full-slot, SCS=30 kHz, $BW$=20 MHz).

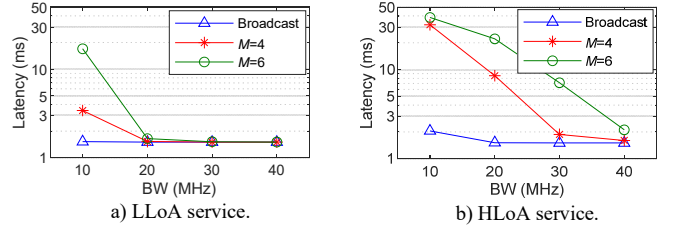

Fig. 5. Average V2N2V $\overline{l_{radio}}$ as a function of the $BW$ with broadcast and unicast DL (full-slot, SCS=30 kHz, 40 veh/km/lane, $T_p$=100 ms).

Fig. 3 shows that the average $\overline{l_{radio}}$ increases with the number of unicast DL transmissions ($M$) for LLoA and HLoA. This is due to an increase in the usage of RBs with $M$ that also augments the time $t_w$ UEs must wait for resources in the DL and hence $\overline{l_{radio}}$. This effect also explains the increase in the percentage of packets dropped with $M$[4]. It is also important to note that the average $\overline{l_{radio}}$ for HLoA in Fig. 3 is already higher than the radio latency requirement established by 3GPP (i.e. 6 ms) when $M$=6 and $T_p$=20 ms for all evaluated densities, as well as for densities equal to or higher than 40 veh/km/lane when $T_p$=100 ms. In these cases, it will not be possible to satisfy the reliability requirements. When the traffic density is equal to 40 veh/km/lane and $T_p$=100 ms, it is necessary a $BW$ equal to 30 and 40 MHz when $M$=4 and $M$=6, respectively, to guarantee an average $\overline{l_{radio}}$ lower than the latency requirement for HLoA. When $T_p$=20 ms, $\overline{l_{radio}}$ is still above 6 ms with $BW$=50 MHz. These results highlight the challenge to scale V2X services with V2N2V communications using unicast DL transmissions.

### 2) Broadcast DL transmissions

The previous section has shown that unicast DL transmissions can compromise the support of V2X services with stringent latency and reliability requirements. This sub-section and the following ones consider broadcast DL transmissions.

Fig. 4 represents the average V2N2V radio latency $\overline{l_{radio}}$ as a function of the traffic density when using broadcast DL transmissions without retransmissions (Section III.A). Fig. 4.b shows that it is possible to guarantee an average $\overline{l_{radio}}$ between 1.5 and 1.58 ms when $T_p$=100 ms for LLoA and HLoA and all the evaluated densities. These results are in line with the results in [3] and [4] that report a radio latency equal to 1.84 ms when using 30 kHz SCS and full-slot transmissions; we should note that [3] and [4] consider a single UE in the scenario. Average $\overline{l_{radio}}$ values of approximately 1.5 ms can also be guaranteed for the LLoA service for densities equal to or below 60 veh/km/lane when $T_p$ decreases to 20 ms (Fig. 4.a). On the other hand, the HLoA service experiences significantly higher average $\overline{l_{radio}}$ values when the traffic density increases and $T_p$=20ms. This is again due to the increased occupancy of the RBs resulting from the use of HEP MCSs with higher error protection. We should though note that under all scenarios, the average $\overline{l_{radio}}$ values are significantly lower for both LLoA and HLoA when using broadcast DL transmissions (Fig. 4) compared to unicast DL transmissions (Fig. 3). This is due to a more efficient use of RBs: broadcast DL transmissions reduce the percentage of used RBs by up to 72% and 82% compared to when using unicast DL transmissions with $M$ equal to 4 and 6 respectively.

Fig. 5 shows the average V2N2V radio latency $\overline{l_{radio}}$ for different values of $BW$ when using broadcast[5] and unicast ($M$=4 and $M$=6) DL transmissions. The figure shows the results for the LLoA and HLoA services. Fig. 5 shows that the LLoA service can maintain an average $\overline{l_{radio}}$ of approximately 1.5 ms with only 10 MHz bandwidth when using broadcast DL transmissions. Using unicast DL transmissions requires increasing the bandwidth to 20 MHz to guarantee $\overline{l_{radio}}$ values between 1.5 and 1.65 ms. With 10 MHz bandwidth, the average $\overline{l_{radio}}$ increases

---

[4] $\overline{l_{radio}}$ in Fig. 3.a when $T_p$=20 ms and traffic density is equal to 80 veh/km/lane is similar for both values of $M$. This is because $\overline{l_{radio}}$ is calculated considering only the packets that are transmitted (and not dropped). Although the percentage of transmitted packets is lower when $M$=6 compared to when $M$=4,

the higher number of unicast DL transmissions performed per packet results in a similar use of RBs and then in similar average $\overline{l_{radio}}$ latencies.

[5] This figure focuses on the average radio latency for the user plane, and hence the results achieved for broadcast transmissions are also valid for multicast transmissions.

to 3.4 and 17 ms when $M$ equal to 4 and 6, respectively. Similarly, the HLoA service can maintain $\overline{l_{radio}}$ values below 2 ms with $BW$=10 MHz using broadcast DL transmissions, and the minimum bandwidth necessary to guarantee an average $\overline{l_{radio}}$ below the HLoA latency requirement (6 ms) increases to 30 and 40 MHz using unicast DL transmissions with $M$ equal to 4 and 6, respectively. These results clearly highlight the efficiency of broadcast DL communications, and their relevance for scalably supporting advanced V2X services. The same trends in terms of the scalability of broadcast compared with unicast DL transmissions are observed for $T_p$= 20 ms.

### 3) k-repetitions or HARQ retransmissions

The HLoA service demands a reliability of 99.99% and has hence been configured with HEP MCSs. HEP MCSs adapt the MCS to guarantee a correct reception of 99.999% of the packets. LEP MCSs adapt the MCS to guarantee the correct reception of 90% of the packets, but the reliability can be improved by transmitting $k$ repetitions of a packet or using HARQ. To evaluate the impact of using $k$-repetitions and HARQ with LEP MCSs, this section focuses on the HLoA service requirements, and compares the latency performance that can be achieved with HEP MCSs, and with LEP MCSs using $k$-repetitions (Section III.C) and HARQ with a maximum of $n$ retransmissions (Section III.D). HEP MCSs and LEP MCSs with $k$-repetitions can be applied with broadcast and multicast transmissions, while LEP MCSs HARQ is considered with multicast transmissions. The reliability is defined as the percentage of packets that satisfy the latency requirement. The reliability depends on the probability of correctly receiving a packet (that depends on the BLER) and the probability that the packet will be received before the deadline. The maximum reliability that can be achieved is then bounded by the probability of correctly receiving a packet and is calculated as $P_{rel} \leq$ 1-BLER$^k$ and $P_{rel} \leq$ 1-BLER$^{n+1}$ when $k$-repetitions and HARQ with a maximum of $n$ retransmissions are used, respectively, with BLER representing the BLER experienced by each (re-)transmission or repetition of a packet. In this context, we consider $k$=4 and $n$=3 in order to increase the reliability to 99.99%. Repetitions of a packet are scheduled in DL with SPS and in UL with Configured Grant, while HARQ retransmissions are scheduled with dynamic scheduling since we cannot anticipate if retransmissions are necessary.

Fig. 6 shows the average V2N2V $\overline{l_{radio}}$ for different values of traffic density using LEP MCSs with $k$-repetitions and HARQ retransmissions and using LEP MCSs (without repetitions or retransmissions). Fig. 6 shows that LEP MCSs with $k$-repetitions result in the highest $\overline{l_{radio}}$ values ($\overline{l_{radio}}$ is always above 4.5 ms) due to the highest use of RBs. As a result, $\overline{l_{radio}}$ with LEP MCSs and $k$-repetitions increases rapidly with the traffic density when $T_p$=20 ms; $\overline{l_{radio}}$ is higher than the 6 ms latency requirement for HLoA for traffic densities higher than 10 veh/km/lane. The use of HEP MCSs results in the lowest latency in the scenarios with lower load (when $T_p$=100 ms or when $T_p$=20 ms and the traffic density $\leq$ 20 veh/km/lane). The latency experienced with HEP MCSs increases significantly in the scenarios with higher network load, i.e., when $T_p$=20 ms and the traffic density is equal to or higher than 40 veh/km/lane. This is due to the congestion of the cell. The lowest $\overline{l_{radio}}$ as the density increases is achieved with LEP MCSs and HARQ retransmissions (Fig. 6). This is due to a more efficient use of RBs since retransmissions only take place when previous packets have not been correctly received. The efficient use of RBs results in that LEP MCSs with HARQ retransmissions can maintain nearly the same $\overline{l_{radio}}$ values for almost all the traffic densities and $T_p$ evaluated, which is not the case for the other service configurations.

### 4) Numerologies, full-slots and mini-slots

Previous results were obtained using full-slots and an SCS of 30 kHz. Latency can be reduced using a higher numerology (i.e., higher SCS) and/or mini-slots. Fig. 7 represents the average V2N2V radio latency $\overline{l_{radio}}$ and the percentage of used RBs when using full-slots and mini-slots (with 7 and 4 OFDM symbols, OS) with different SCSs. Results in Fig. 7 are obtained considering that packets are transmitted only once and there are no retransmissions (Section III.A). Fig. 7.a shows that increasing the SCS reduces the average $\overline{l_{radio}}$ with the lower traffic densities ($\leq$40 veh/km/lane). However, the use of the 60 kHz SCS results in higher $\overline{l_{radio}}$ values when the density increases. This is the case because 60 kHz SCS uses ECP while 30 kHz SCS uses NCP. With ECP, there are only 12 OS per slot to transmit data while with NCP there are 14 OS. LLoA (with LEP MCSs) can then transmit 300-byte packets using on average 2.46 RBs and 2.82 RBs when NCP and ECP are used, respectively. 60 kHz SCS and ECP increase then the use of the RBs (Fig. 7.b) compared to 30 kHz SCS and NCP. When the density increases to 60 veh/km/lane or more, 60 kHz SCS and ECP congest the cell as more than 98% of RBs are utilized (Fig. 7.b), and this increases the average $\overline{l_{radio}}$ (Fig. 7.a).

Fig. 7.a shows that the use of full-slots with higher numerologies (60 kHz SCS) can achieve similar latency values than using mini-slots with lower numerologies (30 kHz SCS with 7OS mini-slot in Fig. 7.a) under low and medium traffic

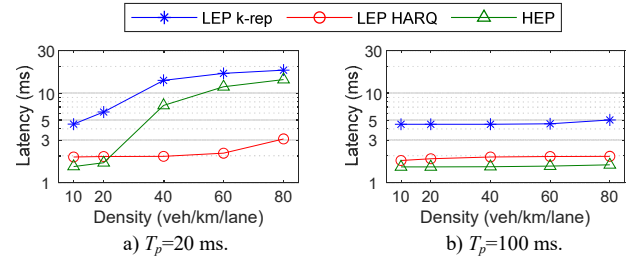

Fig. 6. Avg. V2N2V $\overline{l_{radio}}$ as a function of the density for LEP MCSs with $k$-repetitions and HARQ, and HEP MCSs (full-slots, SCS=30 kHz, $BW$=20 MHz).

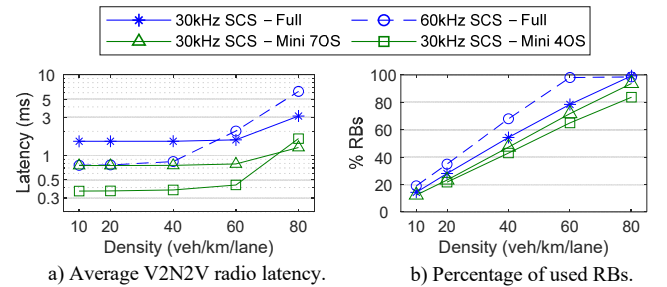

Fig. 7. LLoA performance as a function of the traffic density using full- and mini-slots ($BW$=20 MHz, $T_p$=20 ms).

densities. However, the use of mini-slots reduces the average $\overline{l_{radio}}$ compared to full-slots when the density increases (>40 veh/km/lane) due to the lower use of RBs in the cell (Fig. 7.b). Fig. 7.a also shows that the average $\overline{l_{radio}}$ for the 30 kHz SCS can be reduced by more than 50% and 75% when using 7OS and 4OS mini-slots instead of full-slots under low and medium traffic densities[6]. However, the latency gains obtained with mini-slots decrease when the traffic density increases (Fig. 7.a).

### 5) Support of LLoA and HLoA cooperative lane change

The previous sections have analyzed the impact of the 5G radio configuration on the average latency. This section now evaluates under which configurations can 5G NR satisfy the latency requirements of the cooperative lane change service with LLoA and HLoA. As discussed in Section IV, 3GPP establishes a maximum UL+DL radio latency of 23 ms and a reliability of 90% for LLoA. HLoA requires a UL+DL radio latency below 6 ms and a reliability of 99.99%. The reliability is defined as the percentage of packets that satisfy the latency requirement. This section considers broadcast and multicast in DL given the challenge to scale V2X services with unicast DL transmissions following the results in Section V.A.2.

Fig. 8 shows the average V2N2V (or UL+DL) radio latency $\overline{l_{radio}}$ and the maximum $l_{radio}$ experienced by 90% of the packets for LLoA as a function of the traffic density. Fig. 8 does not consider the use of $k$-repetitions or HARQ retransmissions. Results are shown for different SCS and using full- and mini-slots with 7OS[7]. Results are reported for $T_p$ =20 ms and $BW$=20 MHz since significantly lower latency values are obtained with $T_p$=100 ms or $BW$=40 MHz[8]. Fig. 8 shows that both the average and maximum latency experienced by 90% of the packets is below 6 ms in all the evaluated scenarios. The LLoA latency and reliability requirements established by the 3GPP are hence satisfied with LEP MCSs transmitting a packet only once, and $k$-repetitions or HARQ are not necessary. We should note that the results in Fig. 8 for low densities are similar to the radio latency values reported in [3] and [4] for a single UE. For example, [3] reports radio latency values equal to 1.95, 1.06, and 0.7 ms for 7OS mini-slots and 15, 30 and 60 kHz SCS respectively, while the maximum $l_{radio}$ experienced by 90% of the packets for 20 veh/km/lane in Fig. 8.b is equal to 2, 1.01, and 0.83 ms, respectively. Fig. 8 shows that the latency increases with the traffic density, and hence it is necessary to study the scalability of V2N2V communications.

HLoA has more stringent requirements than LLoA. In particular, HLoA requires a reliability of 99.99% and a radio latency of 6 ms. This reliability can be met with HEP MCSs when packets are transmitted only once, or with LEP MCSs using $k$-repetitions or HARQ retransmissions (Section V.A.3). We only consider HARQ retransmissions with LEP MCSs and multicast transmissions since $k$-repetitions considerably increases the usage of RBs compared to HARQ and can hence compromise the scalability of the network. Fig. 9 shows the percentage of packets that experienced a V2N2V radio latency $l_{radio}$ lower than 6 ms with HEP MCSs and with LEP MCSs and HARQ retransmissions with $n$=3. Fig. 9.a shows that it is possible to guarantee a latency lower than 6 ms for 99.99% of the packets with HEP MCSs when $T_p$=100 ms and the SCS is equal to 15 and 30 kHz using both full-slot and mini-slots; mini-slots in Fig. 9 are configured with 7OS. With a 60 kHz SCS, the percentage of packets that experienced a latency lower than 6 ms decreases to 90.9% and 81.8% when using full-slots and mini-slots, respectively. This is due to the higher usage of RBs with the 60 kHz SCS (Fig. 7.b) that increases the time UEs wait for allocated resources ($t_w$ in (4) and (5)), and hence the latency. If $BW$ increases to 40 MHz, the percentage of packets that experienced a $l_{radio}$ lower than 6 ms augments to 99.99% with HEP MCSs and 60 kHz SCS (full- or mini-slots) since $t_w$ is reduced thanks to the higher availability of RBs. A different trend is observed with LEP MCSs and HARQ retransmissions. In this case, Fig. 9.b shows that 99.99% of the packets are

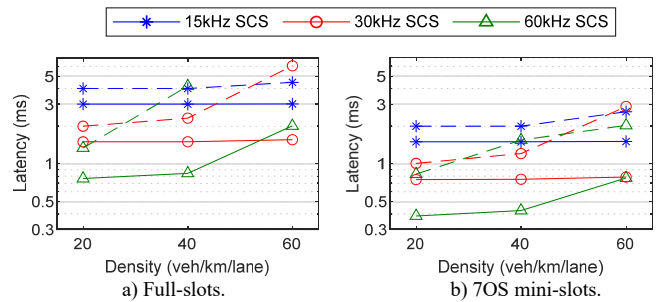

Fig. 8. Average (solid lines) and maximum $l_{radio}$ experienced by 90% of the LLoA packets (dashed lines) ($BW$=20 MHz, $T_p$=20 ms).

a) Full-slots.  b) 7OS mini-slots.

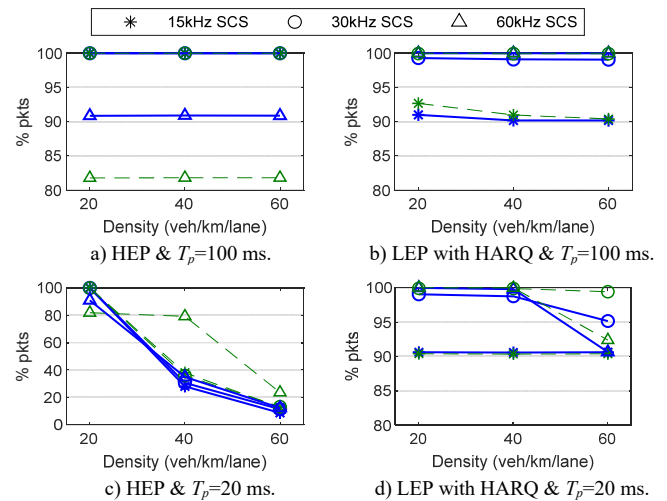

a) HEP & $T_p$=100 ms.  b) LEP with HARQ & $T_p$=100 ms.

c) HEP & $T_p$=20 ms.  d) LEP with HARQ & $T_p$=20 ms.

Fig. 9. Percentage of packets that experienced a $l_{radio}$ lower than 6 ms as a function of traffic density ($BW$=20 MHz). Blue solid line represents the results for full-slots, and green dashed line represents the results for mini-slots.

---

[6] These results are in line with [3] and [7] that showed that the use of mini-slots with 7OS and 4OS could reduce the radio latency by approximately 45% and 65%, respectively, compared to full-slots when using a SCS of 30 kHz. The results in [3] and [7] were though obtained for a single UE in the scenario.

[7] When a packet is dropped, we consider that the packet experiences an infinite latency. The maximum value of the latency experienced for 90% of the packets is then infinite when the percentage of dropped packets is higher than

10%. In this case, no value is depicted in Fig. 8. This happens for 60 kHz SCS, full-slots, 60 veh/km/lane and $T_p$=20 ms when $BW$=20 MHz. With $BW$=40 MHz, the maximum latency for 90% of the packets under this radio configuration is 1.2 ms.

[8] The 3GPP latency and reliability requirements for LLoA with $T_p$ =100 ms or $BW$=40 MHz are satisfied for all traffic densities and SCS evaluated. The maximum latency for 90% of packets is always below 4 ms for these scenarios.

transmitted in less than 6 ms with 60 kHz SCS (with full- or mini-slots) as well as with 30 kHz SCS and mini-slots. The percentage of packets transmitted in less than 6 ms decreases to 99.1% with 30 kHz SCS and full-slots, and to 90.2% and 91% with 15 kHz SCS and full- and mini-slots, respectively. This is due to the highest duration of the slots (full or mini) with lower SCSs that increases the transmission time and the frame alignment time ($t_{tt}$ and $t_{fa}$ in (4), (5), (14) and (15)), and hence the latency. This has a high impact when using retransmissions since $t_{tt}$ and $t_{fa}$ affect the initial transmission of a packet and each of its retransmissions. If $BW$ increases to 40 MHz, similar latency values are observed with LEP MCSs and HARQ for all configurations of SCS and slot (full or mini). This is the case because LEP MCSs and HARQ make an efficient use of RBs and the experienced latency is not affected by the amount of RBs available when $T_p$=100 ms and $BW$ equal to 20 and 40 MHz.

The traffic load increases when $T_p$ decreases to 20 ms, and this significantly reduces the percentage of packets that experienced a $l_{radio}$ lower than 6 ms with HEP MCSs for densities equal to or higher than 40 veh/km/lane and all SCSs (Fig. 9.c, please note the change on $y$-axis limits for this subfigure). The degradation observed is due to an increase in the usage of RBs that reduces the availability of RBs and increases the latency. If $BW$ increases (and hence the number of available RBs), the percentage of packets that experienced a $l_{radio}$ lower than 6 ms increases as shown in Fig. 10. This figure depicts the percentage of packets that experienced a $l_{radio}$ lower than 6 ms when $BW$ varies from 20 to 50 MHz for 60 veh/km/lane. Fig. 10.a shows that a $BW$=50 MHz is needed to guarantee that 99.99% of the packets are transmitted in less than 6 ms for all configurations of SCS and slot. The degradation observed when the traffic density increases (and $BW$ is maintained equal to 20 MHz) with $T_p$=20 ms compared to when $T_p$=100 ms is smaller with LEP MCSs and HARQ than with HEP (Fig. 9). In this case, the percentage of packets transmitted in less than 6 ms only decreases when the density is equal to 60 veh/km/lane. The largest degradation is observed with 60 kHz SCS (full- and mini-slots), although the percentage of packets with $l_{radio}$ lower than 6 ms is still above 90%. If $BW$ increases to 30 MHz, the percentage of packets with $l_{radio}$ lower than 6 ms is again equal to the values achieved for lower traffic densities and $BW$=20 MHz (Fig. 10.b).

Fig. 11 represents the average $\overline{l_{radio}}$ and the maximum $l_{radio}$ experienced by 99.99% of the packets with HEP MCSs and with LEP MCSs and HARQ as a function of the traffic density and the SCS (the maximum $l_{radio}$ experienced by 99.99% of the packets is infinite when the percentage of dropped packets is higher than 0.01%, and no value is shown in Fig. 11 in these case). Results in Fig. 11 correspond to the use of mini-slots. Fig. 11.a and Fig. 11.b show that HEP and LEP MCSs with HARQ guarantee an average $\overline{l_{radio}}$ (< 2.2 ms) below the HLoA latency requirement for all SCSs when $T_p$=100 ms. However, the maximum $l_{radio}$ experienced by 99.99% of the packets with LEP MCSs and HARQ using 15 and 30 kHz SCS is much higher than with HEP MCSs. For example, 99.99% of the packets transmitted with HEP MCSs are received in less than 3.5 ms with 60 veh/km/lane and 30 kHz SCS with mini-slots and $T_p$=100 ms. This value increases to 8.5 ms with LEP MCSs and

HARQ. The HLoA reliability requirement cannot be achieved with HEP MCSs and 60 kHz SCS due to the high percentage of packets dropped. When 15 and 30 kHz SCS are used, Fig. 11.a shows that using HEP MCSs satisfies the latency and reliability requirements of HLoA under all densities evaluated. This is not the case for LEP with HARQ that can only satisfy them with the 60 kHz (Fig. 11.b and Fig. 9). When full-slots are used, the experienced $l_{radio}$ increases in comparison with the use of mini-slots. With HEP MCSs, the latency increase is low, and the maximum latency experienced by 99.99% of the packets is still below 6 ms with 15 and 30 kHz SCS for all evaluated densities. This is not the case with LEP MCSs and HARQ. In this case, although the average $\overline{l_{radio}}$ is below the latency requirement for all SCSs and densities, the latency experienced by the 99.99% of the packets is above 6.3 ms (it achieves latency values of 25 ms for 15 kHz SCS). When $T_p$=20 ms, the average $\overline{l_{radio}}$ with HEP MCSs increases with the traffic density even above the 6 ms latency requirement when there are 40 veh/km/lane or more with 15 and 30 kHz SCS, and when there are 60 veh/km/lane with 60 kHz SCS using mini-slots (Fig. 11.c). In addition, the HLoA reliability requirement cannot be satisfied due to the percentage of packets dropped at the transmitter for all SCSs and densities, except for 15 kHz SCS and 20 veh/km/lane. For this last configuration, the reliability requirement is still not satisfied, but in this case, it is due to the maximum latency experienced by 99.99% of the packets that is equal to 7.8 ms.

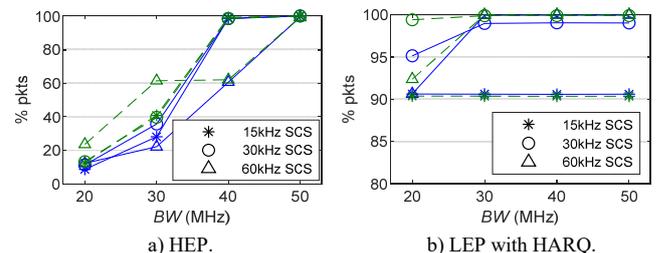

a) HEP.  b) LEP with HARQ.

Fig. 10. Percentage of packets that experienced a $l_{radio}$ lower than 6 ms as a function of $BW$ for 60 veh/km/lane and $T_p$=20 ms. Blue solid lines represent the results for full-slots, and green dashed lines represent the results for mini-slots.

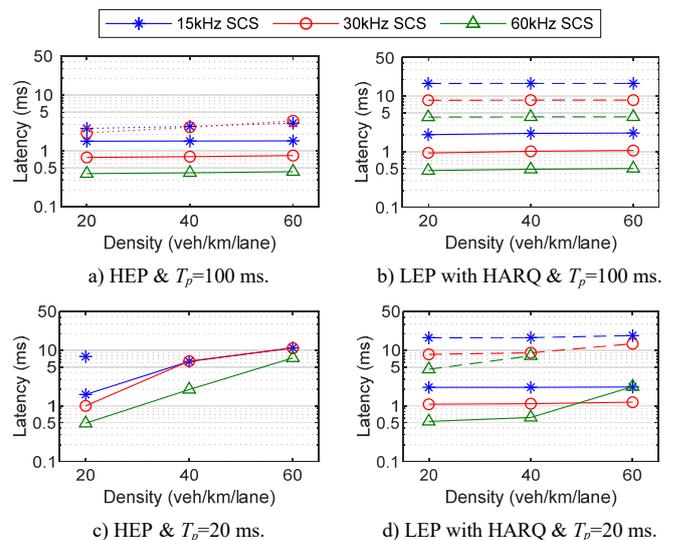

a) HEP & $T_p$=100 ms.  b) LEP with HARQ & $T_p$=100 ms.

c) HEP & $T_p$=20 ms.  d) LEP with HARQ & $T_p$=20 ms.

Fig. 11. Average (solid lines) and maximum $l_{radio}$ experienced by 99.99% of the HLoA packets (dashed lines) ($BW$=20 MHz, mini-slots).

With LEP MCSs and HARQ, $\overline{l_{radio}}$ is maintained nearly equal when $T_p$=20 ms (Fig. 11.d) and $T_p$=100 ms for 15 and 30 kHz SCS (for both SCS, the maximum latency experienced by 99.99% of the packets is above 6 ms). When 60 kHz SCS is used and $T_p$=20 ms, the average $\overline{l_{radio}}$ and the maximum $l_{radio}$ experienced by 99.99% of the packets increase compared with $T_p$=100 ms when the traffic density increases to 40 veh/km/lane or more. In this case, the HLoA reliability and latency requirements are only satisfied when there are 20 veh/km/lane.

### B. *Aperiodic traffic*

This section evaluates the V2N2N radio latency when messages are generated aperiodically.

#### 1) *Dynamic scheduling*

Aperiodic traffic is more efficiently supported using dynamic scheduling. However, dynamic scheduling entails higher latency (see $t_{sch-DS}^{UL}$ and $t_{sch-DS}^{DL}$ in (6) and (7)) since the UE and gNB need to exchange control messages (PDCCH and PUCCH) for each packet in order to request and/or inform about the allocated RBs. The number of RBs reserved for the control messages is important since a low number may queue and delay control messages, and hence impact the radio latency. If the number is too high, it would unnecessarily reduce the number of RBs available for data packets, and hence also impact the radio latency. We evaluate three configurations for the number of RBs reserved for the transmission of control messages. The first one is the baseline configuration (conf.1) in Annex A of [20] (Section IV). The second configuration (conf.2) increases the number of RBs reserved for PDCCH and PUCCH by a factor of 6 and 8 respectively. We consider the third ideal scenario (conf.3) where the control messages can always be transmitted in the next PDCCH or PUCCH after being generated.

Fig. 12 shows the average V2N2V radio latency $\overline{l_{radio}}$ for LLoA with LEP MCSs. Packets are transmitted only once using dynamic scheduling (Section III.B) and DL packets are broadcast[9]. Fig. 12 also shows the average V2N2V radio latency $\overline{l_{radio}}$ for periodic traffic with $T_p$=20 ms. Aperiodic traffic is configured with $T_{avg}$=20 ms, and this results in an average time between packets $\Delta T_a$ of 20 ms. Fig. 12[10] shows that the average latency increases with aperiodic traffic and dynamic scheduling compared to periodic traffic with SPS/Configured Grant even under low densities and considering unlimited capacity to transmit control messages (conf.3). The difference under low densities results from the latency introduced by the exchange of control messages with dynamic scheduling. The difference significantly increases with the traffic density.

Fig. 12 also shows that the number of RBs reserved for control messages impacts the average $\overline{l_{radio}}$. In particular, it can significantly increase the latency if RBs for control messages are under-dimensioned (conf.1 vs conf. 2 in Fig. 12). This also impacts the percentage of packets dropped since the transmitter is still waiting for RBs to transmit control messages by the time a new packet is generated. For example, 60% of the packets are dropped with conf.1 for densities equal to or higher than 20 veh/km/lane. On the other hand, conf.2 does not drop any packets until the density increases to 60 veh/km/lane (i.e., it has sufficient control RBs to manage all generated packets).

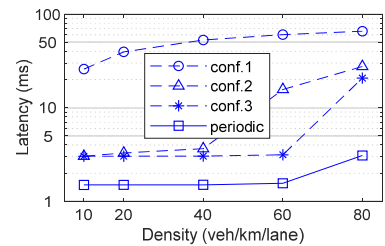

Fig. 12. Average V2N2V $\overline{l_{radio}}$ as a function of the traffic density ($BW$=20 MHz, SCS=30 kHz, full-slots).

#### 2) *Support of LLoA and HLoA cooperative lane change*

We now evaluate if 5G V2N2V communications can support the LLoA and HLoA cooperative lane change service with aperiodic V2X messages and dynamic scheduling. We focus on single broadcast transmissions per packet (i.e. no $k$-repetitions or HARQ), and conf.2 for the selection of the number of RBs reserved for control messages.

Fig. 13 depicts the average V2N2V (or UL+DL) radio latency $\overline{l_{radio}}$ and the maximum $l_{radio}$ experienced by 90% of the LLoA packets (with LEP MCS) when using full- and mini-slots with 7OS. We should first note that the maximum $l_{radio}$ experienced by 90% of the packets is not shown for 60 veh/km/lane (all SCSs), and with densities higher than 40 veh/km/lane for 15 kHz SCS (full- and mini-slots). This is the case because packets dropped at the transmitter prevent receiving more than 90% of the packets. For example, the percentage of packets dropped with 15, 30 and 60 kHz SCS and a density of 60 veh/km/lane is equal to 61.6%, 24.6% and 11.1% for full-slots and equal to 60.3%, 23.2% and 10.2% for mini-slots. Lower SCSs result in higher percentages of packets dropped because the time spent waiting for RBs to transmit the control messages ($t_q^{PDCCH}$ in (8) and $t_w^{PUCCH}$ in (10)) increases with the duration of slots. Fig. 13 shows that it is not possible to guarantee the reliability and latency requirements established by the 3GPP for LLoA ($P_{ref}$=90% and maximum $l_{radio}$ of 23 ms) with 15 kHz SCS for a density equal to or higher than 40 veh/km/lane, or with 30 kHz and 60 kHz SCS for a density equal to or higher than 60 veh/km/lane (using full- or mini-slots). For all the other combinations of SCSs and densities, Fig. 13 shows that the maximum $l_{radio}$ experienced by 90% of the packets is lower than 6.8 and 4.8 ms with full-slots and mini-slots, respectively, which is below the 23 ms requirement. Results in Fig. 13 also show that the use of mini-slots reduces the average $\overline{l_{radio}}$ compared with full-slots thanks to a shorter transmission time ($t_{tr}$ in (4) and (5)). Results in Fig. 13 correspond to $T_{avg}$=20 ms. The conducted analysis for $T_{avg}$=100 ms showed that no packets were dropped at the transmitter for all configurations of SCSs and slots analyzed. In this case, the LLoA reliability and latency requirements were always satisfied, and the maximum $l_{radio}$ experienced by 90% of the packets is lower than 9.6 ms.

---

[9] The impact of using HARQ, $k$-repetitions, different SCSs, full or mini-slots, or unicast DL transmissions with dynamic scheduling is similar to that discussed with periodic traffic and Configured Grant or SPS.

[10] Our results are in line with those reported in [7]. [5] reports a UL latency of 1.2 ms with Configured Grant and of 2.3 ms with dynamic scheduling for a single UE, 30 kHz and full-slots. In our results, the UL latency for low densities is 0.8 ms with Configured Grant and 1.8 ms with dynamic scheduling.

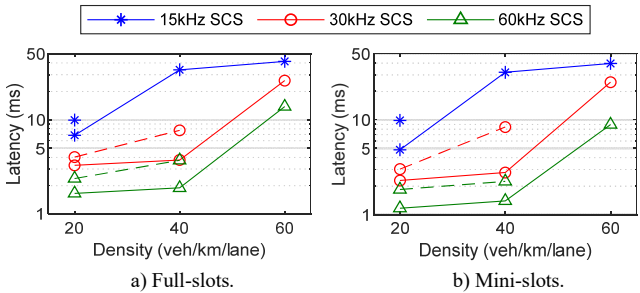

Fig. 13. Average (solid lines) and maximum $l_{radio}$ experienced by 90% of LLoA packets (dashed line) with aperiodic traffic ($BW$=20 MHz, $T_{avg}$=20 ms).

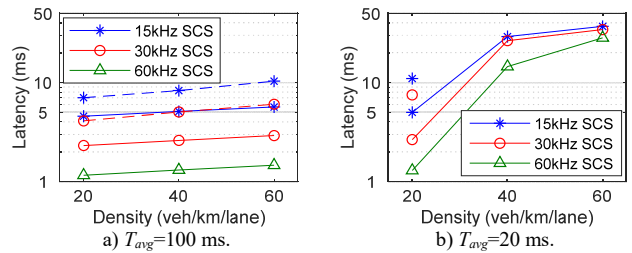

a) $T_{avg}$=100 ms.    b) $T_{avg}$=20 ms.

Fig. 14. Average (solid lines) and maximum $l_{radio}$ experienced by 99.99% of the packets (dashed lines) for HLoA with aperiodic traffic and dynamic scheduling as a function of the traffic density (HEP MCS, $BW$=20 MHz, mini-slots).

We now evaluate the performance of the HLoA service when the traffic is aperiodic and when we use dynamic scheduling. In this case, we focus on the use of HEP MCSs. We do not consider LEP MCSs with HARQ since Fig. 9 showed that with this configuration the HLoA reliability and latency requirements were only satisfied in a few cases with periodic traffic. Since dynamic scheduling introduces a higher latency due to the exchange of control messages (at least equal to 1.5 ms following Fig. 11), LEP MCSs with HARQ would not be able to satisfy the HLoA service requirements with aperiodic traffic. Fig. 14 depicts the average $\overline{l_{radio}}$ and maximum $l_{radio}$ experienced by 99.99% of the packets for the HLoA service with HEP MCSs and mini-slots as a function of the traffic density. Fig. 14 shows that the average $\overline{l_{radio}}$ is lower than 6 ms for all SCSs when $T_{avg}$=100 ms. However, the 15 kHz SCS cannot satisfy the reliability and latency requirements of the HLoA service (i.e. the maximum $l_{radio}$ experienced by 99.99% of packets is higher than 6 ms) for all densities. Percentile results are not shown with 60 kHz SCS since less than 99.99% of the packets are received due to the packets dropped at the transmitter. The HLoA requirements are only satisfied with 30 kHz SCS. Fig. 14 shows that $l_{radio}$ increases considerably for densities of 40 veh/km/lane or more with $T_{avg}$=20 ms. In this case, the maximum $l_{radio}$ experienced by 99.99% of the HLoA packets is always above 6 ms; percentile values are not shown for those configurations for which less than 99.99% of the packets were received. Fig. 13 showed that using full-slots increases $l_{radio}$. With full-slots, the maximum $l_{radio}$ experienced by 99.99% of the HLoA packets increases to 5.0, 5.5 and 6.9 ms for 30 kHz SCS and densities of 20, 40 and 60 veh/km/lane, respectively, when $T_{avg}$=100 ms. The comparison of Fig. 14 and Fig. 11 shows that $l_{radio}$ is higher for all SCS with aperiodic traffic than with periodic.

Fig. 15 shows the maximum $l_{radio}$ experienced by 99.99% of the packets for the HLoA service with aperiodic traffic as a function of the bandwidth $BW$. Fig. 15.a shows that a $BW$ equal to or higher than 30 and 40 MHz is necessary to satisfy the HLoA requirements using full-slots and mini-slots, respectively, when $T_{avg}$=100 ms and there are 60 veh/km/lane. We should note that it is not possible to meet the 6 ms latency requirement with 15 kHz SCS even with a $BW$ of 50 MHz. This is due to the higher slot duration that affects $t_{tt}$ in (4) and (5), $t_q^{PDCCH}$ in (8) and $t_w^{PUCCH}$ in (10). Decreasing $T_{avg}$ to 20 ms increases the cell load.

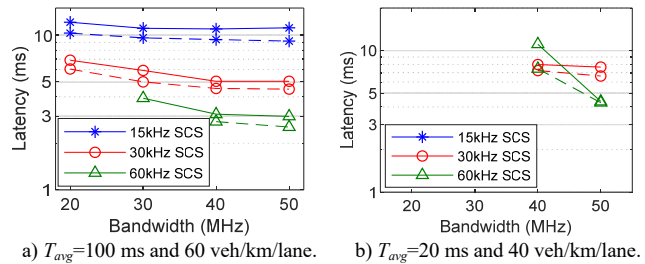

a) $T_{avg}$=100 ms and 60 veh/km/lane.    b) $T_{avg}$=20 ms and 40 veh/km/lane.

Fig. 15. Maximum $l_{radio}$ experienced by 99.99% of the HLoA packets (HEP MCS) with aperiodic traffic and dynamic scheduling as a function of $BW$. Solid lines represent results for full-slots, and dashed lines results for mini-slots.

In this case, the HLoA requirements cannot be satisfied with any of the SCS and $BW$ evaluated with 60 veh/km/lane. If the density decreases to 40 veh/km/lane, Fig. 15.b shows that it is possible to meet the HLoA latency and reliability requirements with 60 kHz SCS and a $BW$ of 50 MHz using both full- and mini-slots.

## VI. DISCUSSION & CONCLUSIONS

This paper has presented an analytical model that estimates the latency experienced in 5G radio networks. The model accounts for a significant number of 5G NR features including the use of different numerologies (SCS, slot durations and CPs), MCSs, the use of full-slots or mini-slots, semi-static and dynamic scheduling, different retransmission mechanisms, and the use of broadcast or unicast transmissions. The model has been used to analyze the capacity to support connected and automated driving services (generally relying on V2V communications) using 5G V2N2V communications, and the impact of different 5G NR radio configurations on the latency. We considered requirements for high and low levels of automation as provided by 3GPP. The implementation of the derived latency models is openly available at [12].

The analysis has highlighted the challenge to scale V2X services with the traffic density when using V2N2V communications with unicast DL transmissions (3GPP Release 16 does not include broadcast/multicast communications). The scalability challenge worsens with the more stringent service requirements that require more reliable MCSs and increase the use of the radio resources. Services can be supported better with V2N2V communications using broadcast or multicast DL communications. In this case, the study showed that 5G can support at the radio network level V2X services with periodic traffic and more relaxed latency and reliability requirements (e.g., cooperative lane change with a low level of automation) using the low error protection MCSs. This is possible even under the highest traffic densities analyzed and is independent of the SCS utilized and regardless of using full- or mini-slots. For all the scenarios evaluated, 5G could guarantee a V2N2V radio latency $l_{radio}$ below 6 ms for 90% of the packets, i.e. well below the 23 ms requirement established by 3GPP. The conducted study has also shown that it is not possible to satisfy more stringent latency and reliability requirements (i.e. those corresponding to the high level of automation for the cooperative lane change service) with the LEP MCSs and a single transmission per packet. To satisfy these requirements, it is necessary to use HARQ retransmissions with the LEP MCSs. However, HARQ retransmissions with LEP MCSs requires using high SCS and/or mini-slots to reduce the duration of the slots, and avoiding that packet retransmissions excessively increase the radio latency. An alternative to satisfy the more stringent reliability and latency requirements of the HLoA service is the use of the high error protection MCSs. In this case, it is possible to meet the latency requirements under low and medium densities using low SCSs and mini-slots. However, satisfying these requirements under high traffic densities requires augmenting the bandwidth.

The study has also analyzed the impact of the scheduling. Semi-static schemes are more appropriate for periodic messages while dynamic scheduling improves the spectrum efficiency in the case of aperiodic messages. However, dynamic scheduling increases the latency due to the exchange of control or signalling messages, and this study has shown that the dimensioning of the resources reserved for control messages can strongly impact the latency with dynamic scheduling. The study has also shown that 5G V2N2V communications can only support, at the radio level, V2X services with aperiodic traffic and relaxed latency and reliability requiremens under low and medium traffic densities using high SCSs (with full- and mini-slots). Low SCS cannot satisfy the requirements due to the higher slot duration and the latency introduced by signalling messages. 5G V2N2V communications could support the more strict latency and reliability requirements only when assigned a low bandwidth using 30 kHz SCS under low traffic densities and packet generation rates. Satisfying these requirements using higher SCS requires augmenting the bandwidth considerably.

In summary, the conducted study has shown that 5G NR can support, at the radio network level, advanced V2X services using V2N2V communications under certain conditions. The capacity to do so is highly dependent on the radio configuration, the service requirements, and the traffic load. A key contribution to evaluate these dependencies is the analytical models presented in this paper. These models represent a valuable tool for the community to carefully design and dimension 5G radio networks to support connected and automated driving services.